**Predicting a novel Nonmagnetic Weyl Semimetal $Pt_3Sn_2S_2$ a kagome material – A DFT study**

**Ravi Trivedi[1,2], Seetha Lakshmy[3], Heera T. Nair[4], Alok Shukla[5, *], Brahmananda Chakraborty[6,7,*]**

[1]Department of Physics, Karpagam Academy of Higher Education, Coimbatore 641021, Tamil Nadu, India,

[2]Centre for Computational Physics, Karpagam Academy of Higher Education,Coimbatore 641021, Tamil Nadu, India

[3]Department of Physics, Mahatma Gandhi University, Kottayam, Kerala
[4]Department of Physics, Department of Physics Maharaja Sayajirao University of Baroda, Vadodara, Gujarat- 390002
[5]Department of Physics, Indian Institute of Technology Bombay, Powai, Mumbai 400076, India

[6]High Pressure & Synchroton Radiation Physics Division, Bhabha Atomic Research Centre, Trombay, Mumbai-400085, India

[7]Homi Bhabha National Institute, Mumbai

## Abstract

We present a detailed theoretical study of $Pt_3Sn_2S_2$, a layered Kagome-type material "inspired by recent investigations of $Co_3Sn_2S_2$ reported in Nature Communication (2020) 11:3985, Reviews in Physics (2022) 8:100072" whose physical properties remain largely unexplored. Thermodynamic stability was confirmed via formation energy calculations, while mechanical stability was evaluated using the Voigt–Reuss–Hill approximation and elastic stability conditions. Dynamical and thermal stability were validated through phonon dispersion and ab initio molecular dynamics simulations, with Pugh's criterion classifying the material as ductile. Spin-orbit coupling (SOC) induces band splitting, giving rise to Weyl points ($W_1$, $W_2$) and a Weyl semimetal phase, accompanied by a SOC-driven transition from non-relativistic band touching to relativistic band inversion, marking a topological phase shift. Under applied pressure, the coupling of spin and valley degrees of freedom generates spin-valley intertwined Dirac cones, enabling tunable electronic properties for spintronic and valleytronic applications. Furthermore, Boltzmann transport calculations using AMSET reveal a high Seebeck coefficient (80μV/K with SOC) and low thermal conductivity, highlighting its potential for high-performance thermoelectric devices. These findings establish $Pt_3Sn_2S_2$ as a promising material for applications spanning topological physics, thermoelectrics, and advanced electronic technologies.

## Introduction

Materials with Dirac-cone structures have some attractive characteristics from the point -of-view of both in basic and applied research. The Dirac-cone structure leads to massless fermions in graphene that give rise to QHEs (Quantum Hall Effect) with half-integer, fractional, and fractal scales and ultrahigh carrier mobility, among many other unique phenomena and features **[1]**. Dirac cone defines a linear relationship between energy and momentum, i.e.;the energy v/s momentum relation forms Dirac cones. They are singularities in the spectrum of Hamiltonians and the linear energy dispersion around discrete locations that are topologically protected. They are unique energy bandswith massless electrons, resulting in extremely high electron mobility **[2]**. In condensed matter physics, massless or massive Dirac cones lead to attractive phenomena depending on the dimensionality, spins or the type of materials, including Klien tunnelling, high mobility and novel topological phases **[3-11]**. The kagome materials have attracted much attention,revealing various intriguing electronic states such as Dirac fermions **[12,13]**, and spin-liquid phases **[14–19]**. Recently, a cobalt-based compound $Co_3Sn_2S_2$ arranged in an overlapping triangle was investigated and found that an electronic band connecting the two Weyl cones is flattened due to the electron-correlation effects and emerges near the Fermi energy in $Co_3Sn_2S_2$**[20]**. It has been observed that in the presence of strong SOC, topological insulators, $Sb_2Te_3$, $Bi_2Te_3$ and $Bi_2Se_3$ have robust surface states consisting of a single Dirac cone**[21]**. Furthermore, the emergence of edge states, the existence of Weyl nodes, and charge density waves have been observed in Fe-Sn **[22-23]**, $Co_3Sn_2S_2$**[20]**, $Fe_3Sn_2$**[24]**, and Co-Sn **[25]** Kagome lattice compounds. Recently, compounds $T_3Pb_2Ch_2$(T=Pd,Pt andCh=S,Se) shandite structure have been studied theoretically and found several flat bands around the Fermi level with the existence of Dirac points along theΓ–T direction **[26]**. Pressure is a particularly effective way of tuning the lattice structure and manipulating electronic states such as topological phases of matter **[27-32]**. Under pressure, a large Berry curvature dominating the Hall Effect in $Pb_{1x}Sn_xTe$ material that displayed the topological metallic phase was observed**[27]**. The nontrivial influence of hydrostatic pressure also leads to the appearance of four two-fold degenerate Dirac cones at high pressure in bulk phosphorous, which form unusual nodel rings **[28]**. Pressure-induced superconductivity in the non-centrosymmetric Weyl semimetals LaAlX (X = Si,Ge) has also been reported by Cao et al. [24]. They found that, due to the antisymmetric SOC, novel superconducting properties arise in these materials.

## Computational Methodology

The electronic structure calculations were performed using the DFT-based Vienna Ab initio Simulation Package [33-36], having a plane wave basis set and exchange-correlation functional within the generalized gradient approximation (GGA). The atomic positions and the unit cell dimensions were optimized until the forces were less than 0.01 eV/Å. The lattice parameters a=b=6.08 Å, c=13.33Å perfectly match with the experimental lattice parameters of the $Co_3Sn_2S_2$ kagome type lattice [20]. Spin-orbit coupling (SOC) is also included to drive the topological behaviour and stability. For the simulations including relativistic effects, i.e., with SOC, the atomic coordinates optimized using the scalar non-relativistic effects are re-relaxed. The Brillouin zones (BZ) were sampled using the Monkhorst-Pack scheme [37]with K point grids of 7x5x3 for optimization, and 11x9x5 for accurate calculation of density of states (DOS). A dense 14x14x14 K point grid was also used to explore the Fermi surface, and it was visualized using Xcrysden [38] software. Analysis of thermoelectric parameters like the Seebeck coefficient and thermal conductivity was calculated in the framework of DFT using BoltzTraP2 codes [39] implemented in AMSET [40]. In addition, pressure-induced effects in the system are also studied by varying the lattice constant *a*.

## Results and discussion

$Pt_3Sn_2S_2$, a transition metal-based shandite compound having the space group of R-3m (No. 166), crystallizes in a trigonal/rhombohedral phase similar to that of $Co_3Sn_2S_2$ (experimentally synthesized) type kagome lattice [32], in which a quasi-2D $Pt_3Sn$ layer is sandwiched between the S atoms (see Fig 1.). Table 1 shows the structural parameters, including the lattice constants, Wyckoff positions, and x, y, and z coordinates [supplementary information (SI-I)], optimized computationally both with and without the relativistic effects, i.e., SOC.The other details like space group, BZ region, reciprocal lattice etc. are given in supplementary information (SI-1). The quasi-2D character of the $Pt_3Sn_2S_2$ crystal structure induced a slight elongation in the lattice parameter of the conventional unit cell (with or without SOC).

**TABLE I. DFT optimized parameters of $Pt_3Sn_2S_2$ (space group R-3m) with and without SOC. The conventional unit cell is expressed in angstrom, and cell angles α, β, γ are also presented. [First row values taken from Co3Sn2S2 parameters [32], as we just replaced the Co atom via Pt atom, second and third row after the optimization]**

| | a | b | c | α | β | γ | Wyckoff position of $Pt_3Sn_2S_2$ similar to $Co_3Sn_2S_2$ | | | | |
|---|---|---|---|---|---|---|---|---|---|---|---|
| $Pt_3Sn_2S_2$ | 5.36 | 5.36 | 13.02 | 90 | 90 | 120 | 3a | Sn | 0 | 0 | 0 |
| $Pt_3Sn_2S_2$ Without SOC | 6.08 | 6.08 | 13.31 | 90 | 90 | 120 | 3b | Sn | 2/3 | 1/3 | 5/6 |
| $Pt_3Sn_2S_2$ With SOC | 6.08 | 6.08 | 13.31 | 90 | 90 | 120 | 6c | S | 2/3 | 1/3 | 0.05 |
| | | | | | | | 9d | Pt | 2/3 | 5/6 | 5/6 |

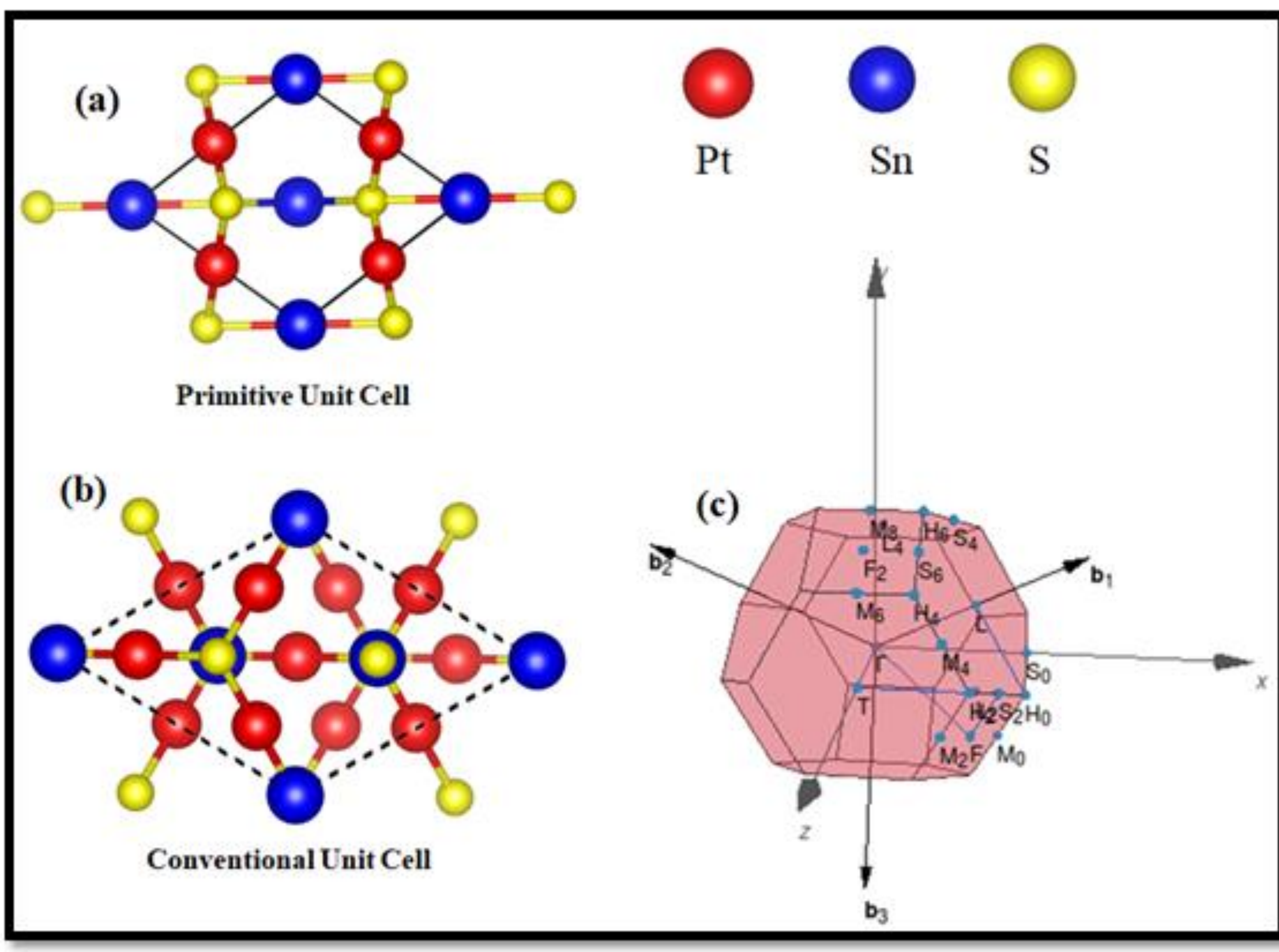

FIG. 1. Crystal structure of $Pt_3Sn_2S_2$in (a) primitive and (b) conventional unit cells. (c) BZ with the location of high symmetry K points of space group R-3m.

Elastic constants, encompassing Bulk modulus (B), Young modulus (E), Shear modulus (G), Poisson ratio (ν), and linear compressibility, were computed to assess the mechanical behavior of the system via second-order derivative matrices.

$$\boldsymbol{C_{ij}} = \frac{\mathbf{1}}{\boldsymbol{V_0}}\left(\frac{\boldsymbol{\partial^2 E}}{\boldsymbol{\partial_{\epsilon_i}\partial_{\epsilon_j}}}\right)$$

In the context, E represents the crystal energy, $V_0$ is the equilibrium volume, and Ɛ denotes strain. The 6x6 elastic matrix, or stiffness matrix, is provided in supplementary information (SI-3). Rhombohedral crystals like Pt3S2S2, with a $\bar{3}$m Laue class and symmetries of 32, 3m, $\bar{3}$m point groups, must adhere strictly to the necessary and sufficient elastic stability conditions outlined by Mouhat F. and Coudert F. **[41]**.

$$\boldsymbol{C_{11}} > |\boldsymbol{C_{12}}|;\ \boldsymbol{C_{44}} > 0$$

$$\boldsymbol{C_{13}^2} < \frac{\mathbf{1}}{\mathbf{2}}\boldsymbol{C_{33}}(\boldsymbol{C_{11}} + \boldsymbol{C_{12}})$$

$$\boldsymbol{C_{14}^2} + \boldsymbol{C_{15}^2} < \frac{\mathbf{1}}{\mathbf{2}}\boldsymbol{C_{44}}(\boldsymbol{C_{11}} - \boldsymbol{C_{12}})$$

The calculated stiffness matrix of $Pt_3S_2S_2$ follows the above condition indicating its mechanical stability.

The *B* and *G* calculations have been carried out using Voigt–Reuss–Hill (VRH) approximation method**[41]**, which is based on the following equations: -

$$\boldsymbol{B_V} = \frac{\boldsymbol{C_{11}+2C_{12}}}{\mathbf{3}},\ \boldsymbol{G_V} = \frac{\boldsymbol{C_{11}- C_{12}+3C_{44}}}{\mathbf{5}},\ \boldsymbol{G_R} = \frac{\boldsymbol{5C_{44}(C_{11}- C_{12})}}{\boldsymbol{4C_{44}+(3C_{11}-C_{12})}}$$

The equation features Voight bulk (BV), Voight shear (GV), and Reuss shear (GR). Values of B, E, G, and ν calculated without SOC are listed in Table 2. Ductility or brittleness is determined by Paugh's criterion: if BH/GH exceeds 1.75, the material is ductile; likewise, if ν surpasses 0.24. $Pt_3S_2S_2$ is ductile, with BH/GH at 9.34 and ν at 0.45, as detailed in supplementary information (SI-4-5).

**TABLE II The calculated elastic constant parameters Voight ($B_V$ ), Reuss ($B_R$ ), average of the Voight and Reuss ($B_H$) for the bulk modulus (in GPa) and, Voight ($G_V$ ), Reuss ($G_R$ ), average of the Voight and Reuss ($G_H$) for the shear modulus, $B_H$ /$G_H$ ratio, Young's modulus E (in GPa), Poisson's ratio (v) for rhombohedral $Pt_3Sn_2S_2$**

| $Pt_3Sn_2S_2$ | $B_V$ | $B_R$ | $B_H$ | $G_V$ | $G_R$ | $G_H$ | $B_H$ /$G_H$ | E | v |
|---|---|---|---|---|---|---|---|---|---|
| 1 | 87.20 | 74.85 | 81.05 | 17.60 | 1.55 | 9.57 | 9.34 | 51.76 | 0.45 |

**Dynamic and thermal stability:**

Phonon dispersion calculations, both with and without SOC have been carried out along the high symmetry K-points of the BZ to confirm the dynamical stability of the $Pt_3Sn_2S_2$ with Fig. 2(a,b) displaying no imaginary frequencies indicating stability. Pt primarily contributes to higher optical modes, while Sn and S dominate middle optical modes. Acoustic phonon branches span frequencies from 0 to 133.6 cm−1 (0 to 4 THz), overlapping with low-frequency optical modes. The less dispersive higher optical modes suggest strong intra-molecular interactions, affirming system stability suggests that this material can be experimentally synthesized.”. In addition to these, ab initio molecular dynamics (AIMD) simulations are performed to examine the thermal stability of the present system. The temperature of the conventional unit cell is raised from 0 to 400 K with a time step of 1 fs by placing it in a microcanonical ensemble for a period of 10 ps. After reaching 400K, the unit cell was transferred to a canonical ensemble with a constant temperature of 400 K. The system attained an equilibrium temperature of 400 K after another 5ps.

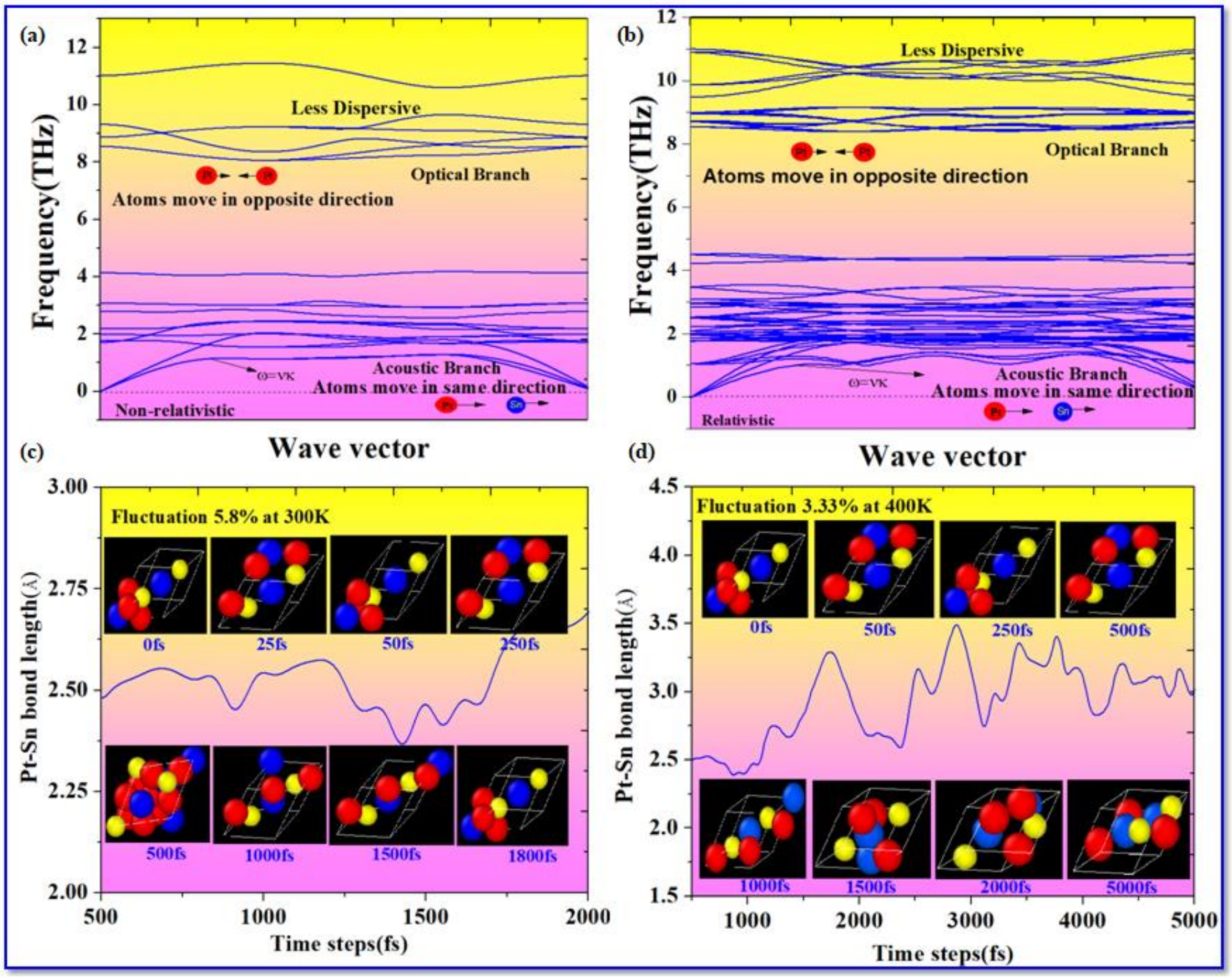

FIG. 2. Dynamic stability (a) for non-relativistic case (b) relativistic case, and thermal stability through AIMD simulation (c) at 300K (b) at 400K [fluctuations are shown with the snapshot at different time steps in fs.

The final structure is analyzed, with MD snapshots in different orientations shown in Figure 2(c,d). Results suggest $Pt_3Sn_2S_2$ maintains its trigonal structure at 300 K, with slight distortions observed at 400 K, though Pt—Sn and Pt—S bonds remain intact. Phonon dispersion and MD simulations indicate the dynamic and thermal stability of $Pt_3Sn_2S_2$. Additionally, thermodynamic stability is confirmed by a formation energy (Ef) value of -0.65 eV/atom, calculated as the difference between the energy of $Pt_3Sn_2S_2$ and the sum of energies of Pt, Sn, and S.

$$E_f = \frac{E_{Pt_ASn_BS_C} - A * E_{Pt} - B * E_{Sn} - C * E_S}{A + B + C}$$

The decomposition of the formation energies to $E_f$/atom (eV) illustrates the contribution of each atom to the formation energy.

**Evolution of electronic states in $Pt_3Sn_2S_2$:**

Fig. 3(a-d) displays the energy dispersion relation, DOS, and band structures, presenting comparisons between results with and without relativistic effects (SOC).

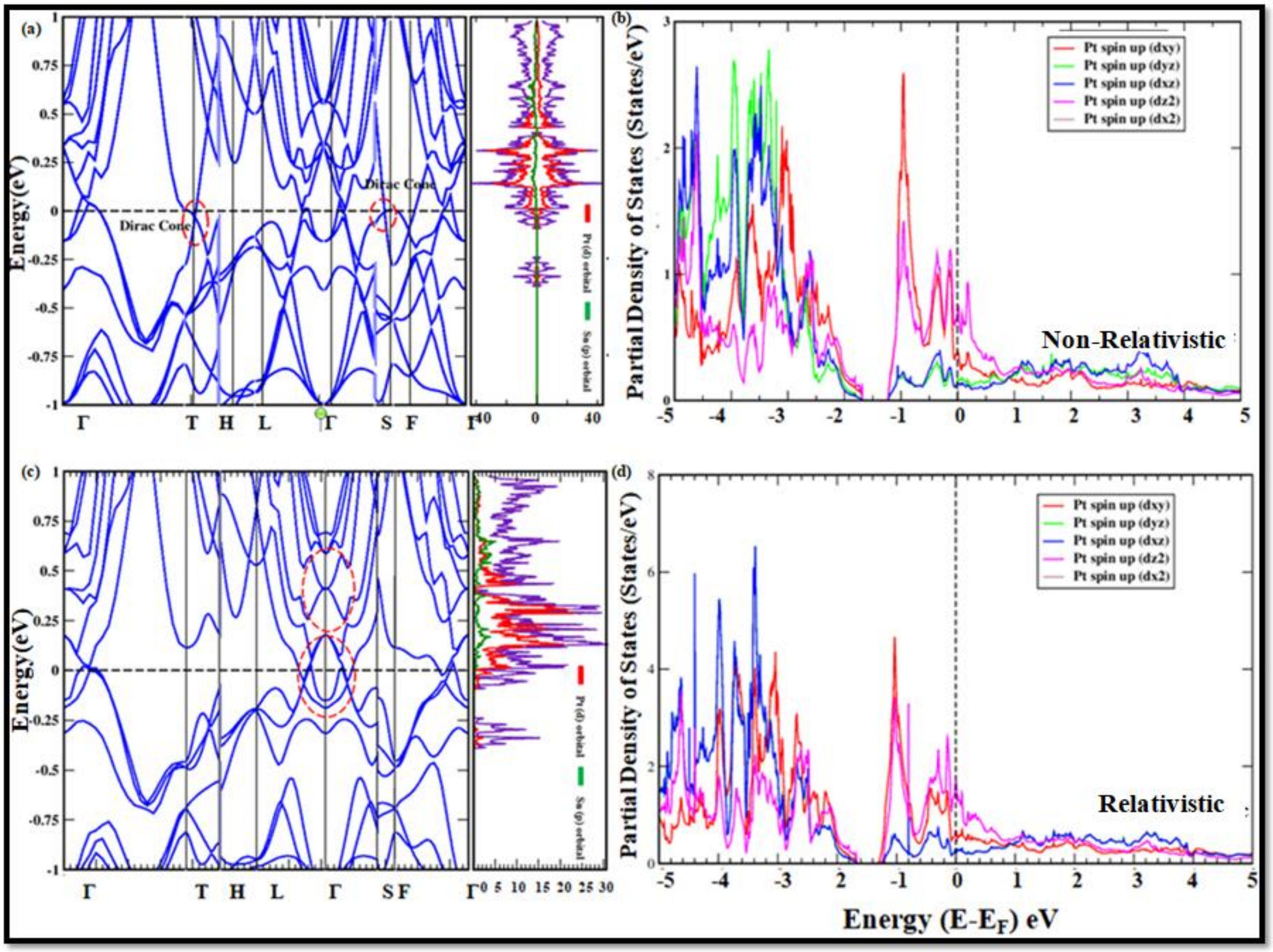

FIG. 3. Comparison of electronic states and energy dispersion relation of $Pt_3Sn_2S_2$ (a,c) without SOC and with SOC. The PDOS illustrated in (b,d). Red circles the Dirac cones in the band structure.

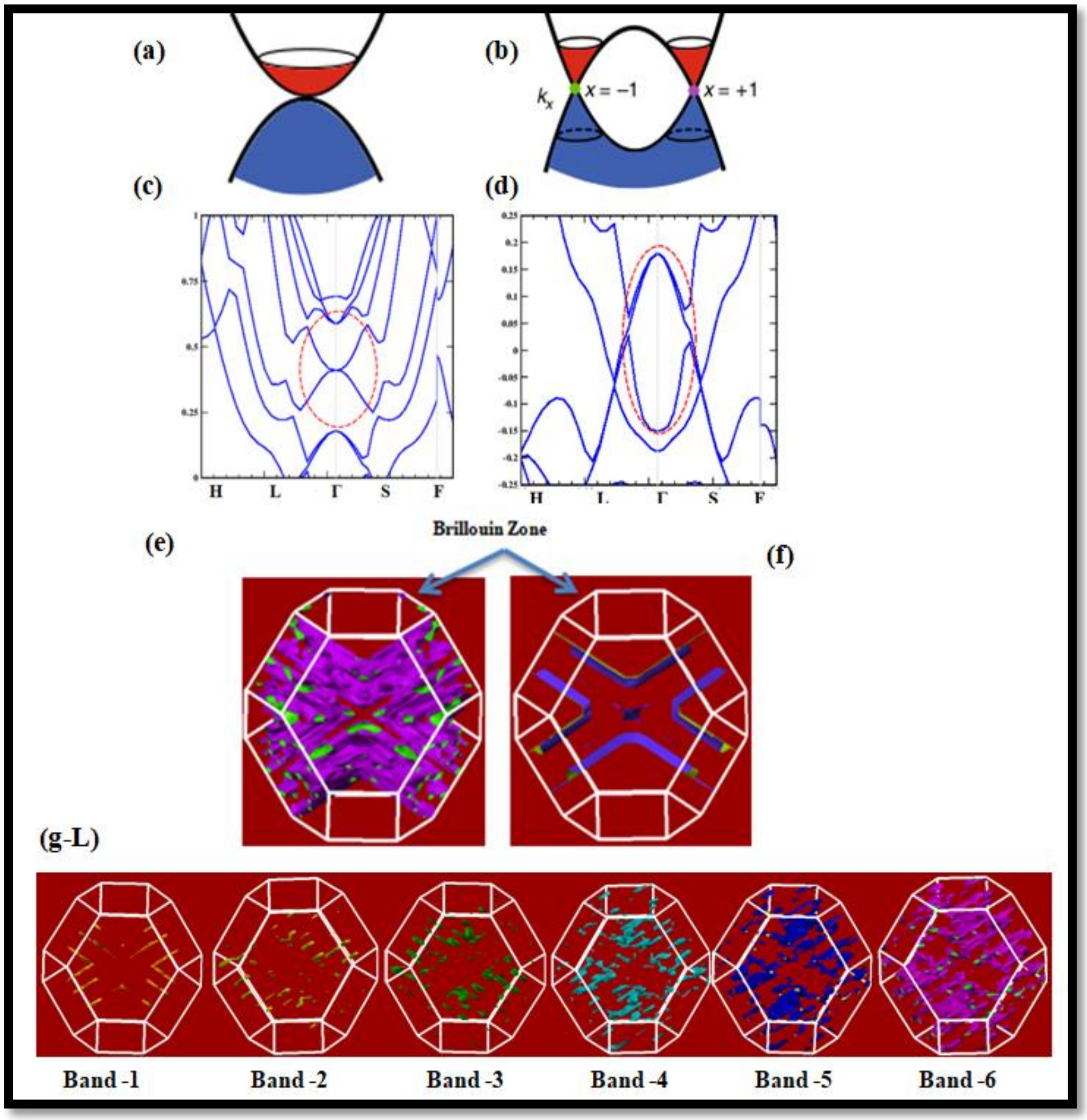

FIG. 4. Schematic representation of energy bands of (a) non relativistic Fermion band touching each other at zero momentum point (b) schematic energy band of a pair of type I Weyl nodes (c) two energy bands of non-relativistic fermion touch each other with SOC at Γ point (d) a pair of type 1 Weyl nodes in the non-relativistic case (e-f) map of Fermi surface in first BZ for both case, i.e. with/without SOC respectively, arising due to all bands near the Fermi level and (g-L) Fermi surface map of different bands for the relativistic case of the electron band crossing. [(a) and (b) have been taken from NATURE COMMUNICATIONS | (2019) 10:1028 with prior permission, http://creativecommons.org/licenses/by/4.0/.]

At the Fermi level, two energy bands linearly merge, forming two zero-gap Dirac cones. In the absence of SOC (see Fig. 3a), the band plot exhibits conical dispersions between T and H, and S and F points along the Γ-x and Γ-y directions (similar to Fig. 1c). $Pt_3Sn_2S_2$ is determined to be non-magnetic, as evidenced by the similar nature of spin-up and spin-down states across all cases. Fig. 3(a,c) illustrates the band structure along high-symmetry K-points (Γ-T-H-L-Γ-S-F-Γ direction). The dispersions primarily correspond to Pt d orbitals, with negligible contribution from Sn orbitals, as indicated by the DOS and PDOS plots in Fig. 3(a,c,b,d). Spin-orbit coupling (SOC) induces spin-splitting in the band structure due to the lack of inversion symmetry, leading to the appearance of Weyl points W1 and W2 **[42-44]**. Also we can notice that at the gamma point, the Dirac cone gets spin splitted, resembling spin-valley intertwined Dirac cone due to strong spin-orbit coupling exhibiting robust Dirac spin-valley coupling. Fig. 4(a) is the sketch of the two parabolic bands for the non relativistic fermions touching at a point, and Fig. 4(b) showing gapless Weyl fermions. Our calculations also predict the same nature of $Pt_3Sn_2S_2$ where Fig. 4(c) exhibits two parabolic bands meeting at the Gamma point because the non-relativistic electrons follow the non-relativistic band dispersion **[42]**. When relativistic effected are included, Weyl femions exhibits band inversion as depicted in Fig. 4(d), which is in line with the reported results for TaAs **[42]**. So for Weyl femions, we can observe a transition from band touching pattern (figure 4c) in case of non-relativistic case to a band inversion pattern (Fig. 4d) in case of relativistic case in both $Pt_3Sn_2S_2$system as well as in TaAs **[42]** system. The Fermi energy is sufficiently close to the Weyl points, the Fermi surface (FS) at the Weyl points drops to zero, as shown in Fig. 4(e-f). Identical behaviour for $TM_3Sn_2$ (TM = Ni, Cu) has recently been reported by S. Baidya et al**.[45]**. When subjected to SOC, the nature of the Fermi surface is different, as shown in Fig. 4(f), which is consistent with broken inversion symmetry.Fermi surfaces of $Pt_3Sn_2S_2$ are also plotted over the first BZ, for the six electron bands as shown in Fig. 4(g-L) crossing $E_F$.

**Pressure-induced effect**

To explore the impact of external pressure on electronic properties, pressure ranging from 5 GPa, 8 GPa, and 20 GPa was applied along the relaxed geometry of $Pt_3Sn_2S_2$. Significant changes in energy dispersion curves were observed as reported in the previous literature on different kagome compounds [27-32]. Fig. 5(a-h) depict band structure, TDOS, and PDOS, plots with and without relativistic effects for pressures of 5 and 20 GPa; supplementary information (SI-6) includes plot for 8GPa pressure. A comparison of Figures 3 and 5 reveals that applying 5 GPa results in the splitting of the Dirac cone and a small band gap (~0.04 eV) opening at T and Γ points. This gap formation may stem from the splitting of the Pt atom's π band into bonding and antibonding π* bands between the degenerate energy band with an energy interval. This phenomenon has been explored in graphene superlattices where π bands split into $\pi_a$ and $\pi_z$ bands with an energy interval $E_s$ to move the Dirac cone **[46]**, showing attractive applications in topological insulators, and superconductivity.

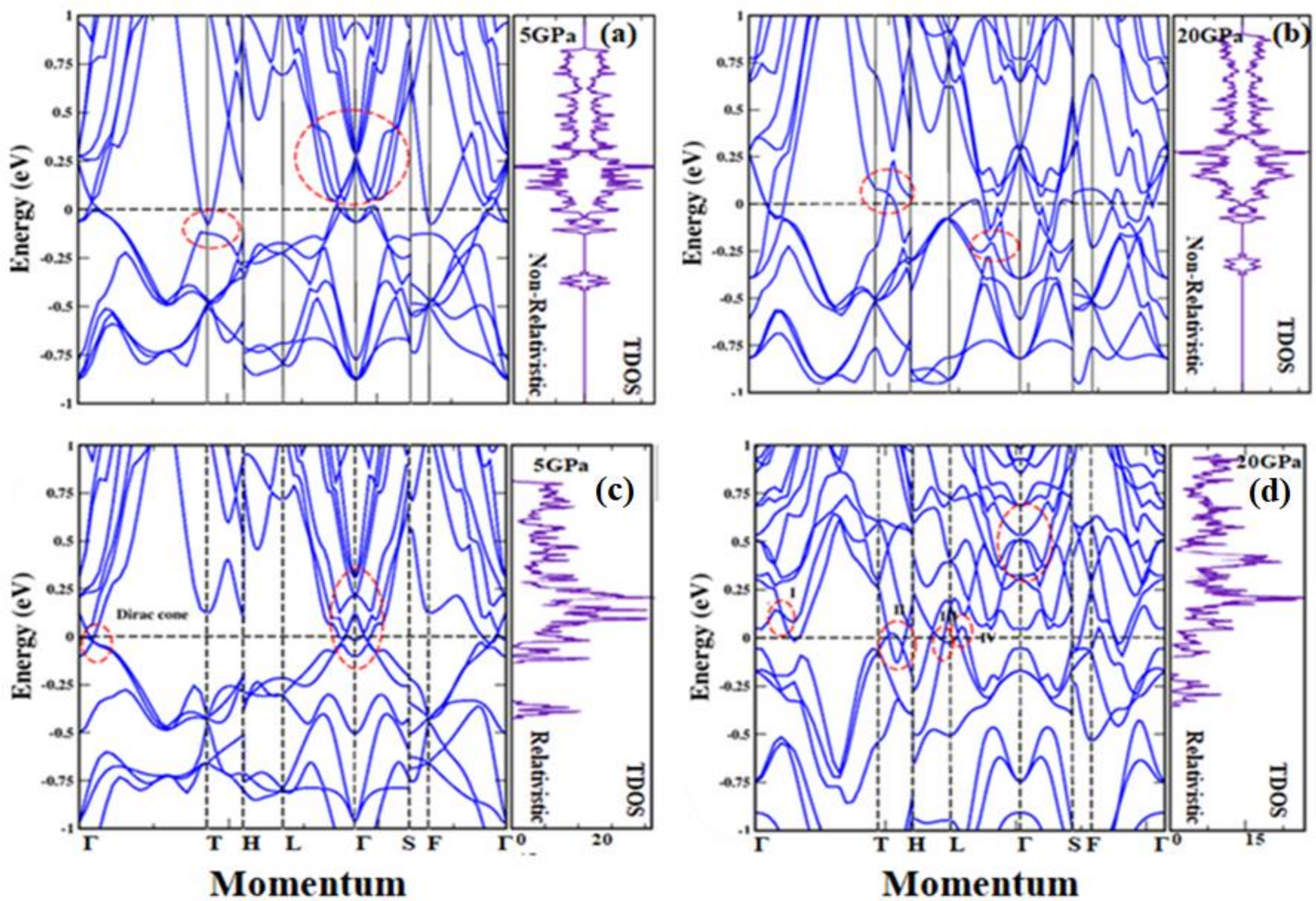

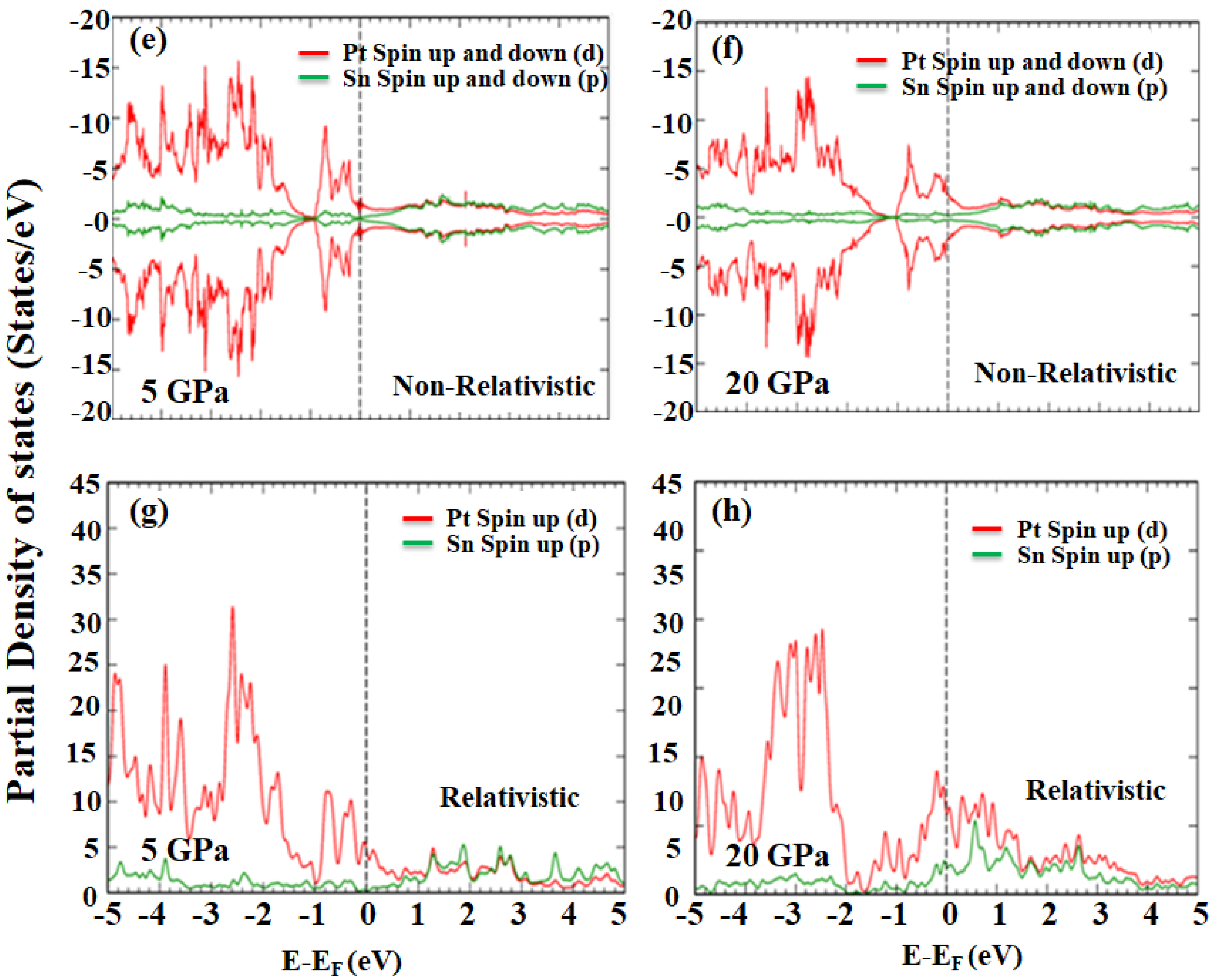

FIG. 5. Electronic band structure along with toal density of states states at 5GPa and 20GPa (a-d) and Partial density of states at (e-h) for non-relativsitic and relativistic case.

In the valence band region, a Dirac cone is identified around 0.25 eV (highlighted in red circle) between F and Γ points, suggesting a shift possibly due to enhanced hopping with $Pt_3Sn_2S_2$ layers and sublattice symmetry breaking [41-42]. Similar shifts in Dirac cones were observed in bilayer sandwich compounds Si2CaSi2 and AA-stacked CaSi2 [47-48]. At higher pressure, such as 20 GPa (Fig. 5d), both parabolic bands touch, increasing DOS at the Fermi level due to enhanced electron pockets. When considering SOC and pressure effects (Fig. 5c, d), a noticeable Dirac cone formation near the Fermi level is observed between Γ-T, T-H, and L-T regions, leading to spin-valley intertwined Dirac cones at the Γ points and threefold charge density waves [CDW] (Fig.5a) due to symmetry breaking and valley-contrasting spin splitting. This phenomenon, observed even without SOC, becomes more prominent with SOC introduction, offering potential applications in spintronics and valleytronics **[49-51]**.

**Thermoelectric properties:-**

Electronic transport properties, namely the Seebeck coefficient and thermal conductivity scattering rates, were calculated using AMSET **[35-36]** code.The Seebeck coefficient is of primary interest in predicting and analyzing thermoelectric response of materials.

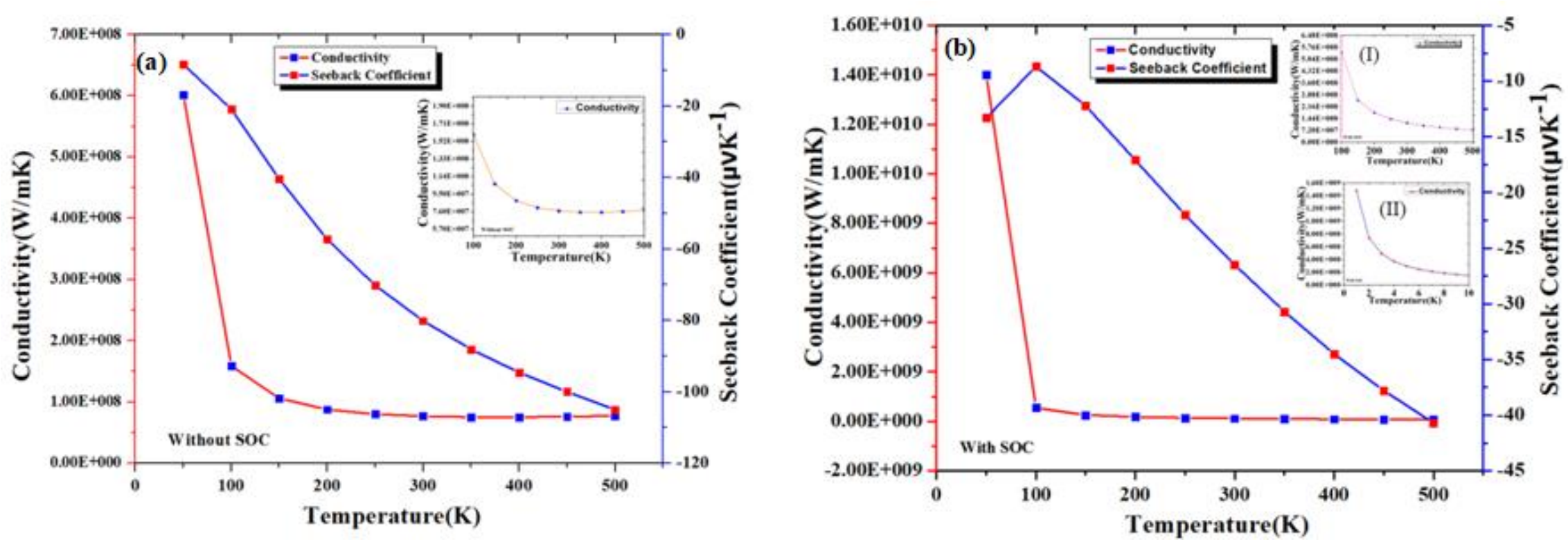

FIG. 6. The variation of conductivity ($10^6$ W/mk) and Seebeck coefficient (μVK$^{-1}$) with temperature (a) without SOC (b) with SOC

The variation indicates in Fig. 6 decrease in the thermal conductivity κ, and increase in the magnitude of Seebeck coefficient S with the increasing temperature, a result consistent with recent findings **[52-53]**. The inverse relationship implies the presence of anharmonic phonon-phonon interactions in the system **[54]**. It is well known that anharmonic phonons lead to a rapid decrease in the thermal conductivity with the increasing temperature, thereby leading to an increase in the magnitude of the Seebeck coefficient. Hence, “anharmonicity” of the lattice vibrations plays a vital role in determining the thermal resistance of a system **[55]** and can be understood in terms of Umklapp scattering **[56]**. It can be observed that the partial electronic DOS shows [**supplementary information SI-7]** that the $s^2$ electrons in the S atom's $3s^2$, valence configuration do not participate in bonding, generating the lone pair configuration and causing severe phonon anharmonicity by interacting with the bonding electrons of nearby Pt atoms **[57]**. At the same time, the large value of the Seebeck coefficient, which is around 80μV/Kand 27.5μV/K with and without SOC, or the low thermal conductivity value, indicates that the material could be useful in building high performance thermoelectric devices **[54]**. The negative values of S indicates that the majority of charge carriers are electrons in the system.

**Conclusions : -**

In summary, we explored the spin-valley intertwined Dirac cone in nonmagnetic Weyl Semimetal $Pt_3Sn_2S_{2.}$ DFT band structure calculations reveal the co-existence of three intertwind charge density waves at gamma points resulting from symmetry breaking leading to valley contrasting spin splitting. Inverse relationship of κ (thermal conductivity) and S (Seebeck coefficient) with temperature owing to the increased phonon scattering due to anharmonic phonon-phonon interaction. This feature of spin-valley intertwined Dirac cone suggests the possibility of using this material in both spintronics and valleytronics.

**Acknowledgements**

RKT, and BC would like to thank Dr. Nandini Garg, Dr. T. Sakuntala, Dr. S.M. Yusuf and Dr. A. K. Mohanty for support and encouragement. RKT, and Brahmananda Chakraborty would also like to thank the staff of BARC computer division for supercomputing facility.

**References: -**